\documentclass{aa}
\usepackage{psfig}

\def\ltsima{$\; \buildrel < \over \sim \;$}
\def\lsim{\lower.5ex\hbox{\ltsima}}
\def\gtsima{$\; \buildrel > \over \sim \;$}
\def\gsim{\lower.5ex\hbox{\gtsima}}

\begin{document}
 
\title{The puzzling case of GRB 990123:\\
multiwavelength afterglow study\thanks{Partly based on observations 
collected at the Bologna Astronomical Observatory in Loiano (Italy) and
and with the William Herschel Telescope of the ING at La Palma (Spain).}}

\titlerunning{The multiwavelength afterglow of GRB 990123}
\authorrunning{Maiorano et al.}

\author{E. Maiorano\inst{1,2}
\and
N. Masetti\inst{1}
\and
E. Palazzi\inst{1}
\and
F. Frontera\inst{1,3}
\and
P. Grandi\inst{1}
\and
E. Pian\inst{4}
\and
L. Amati\inst{1}
\and
L. Nicastro\inst{5}
\and \\
P. Soffitta\inst{6}
\and
C. Guidorzi\inst{7}
\and
R. Landi\inst{1}
\and
E. Montanari\inst{4}
\and
M. Orlandini\inst{1}
\and
A. Corsi\inst{6}
\and
L. Piro\inst{6}
\and
L. A. Antonelli\inst{8}
\and \\
E. Costa\inst{6}
\and
M. Feroci\inst{6}
\and
J. Heise\inst{9}
\and
E. Kuulkers\inst{10}
\and
J.J.M. in 't Zand\inst{9}
}

\institute{
INAF -- Istituto di Astrofisica Spaziale e Fisica Cosmica, Sezione di 
Bologna, Via Gobetti 101, I-40129 Bologna, Italy (formerly IASF/CNR, 
Bologna)
\and
INAF -- Dipartimento di Astronomia, Universit\`a di Bologna, Via Ranzani 
1, I-40126 Bologna, Italy
\and
Dipartimento di Fisica, Universit\`a di Ferrara, Via Paradiso 12, I-44100
Ferrara, Italy
\and
INAF -- Osservatorio Astronomico di Trieste, Via G.B. Tiepolo 11, I-34100
Trieste, Italy
\and
INAF -- Istituto di Astrofisica Spaziale e Fisica Cosmica, Sezione di
Palermo, Via La Malfa 153, I-90146 Palermo, Italy (formerly IASF/CNR,
Palermo)
\and
INAF -- Istituto di Astrofisica Spaziale e Fisica Cosmica, Sezione di
Roma, Via Fosso del Cavaliere 100, I-00133 Roma, Italy (formerly IASF/CNR,
Roma)
\and
Astrophysics Research Institute -- Liverpool John Moores University, 
Twelve Quays House Egerton Wharf Birkenhead CH41 1LD, United Kingdom
\and
INAF -- Osservatorio Astronomico di Roma, via Frascati 33, I-00040
Monte Porzio Catone, Italy
\and
SRON, Sorbonnelaan 2, NL-3584 CA Utrecht, The Netherlands
\and
ESTEC/ESA, SCI-SDG, Keplerlaan 1, NL-2201 AZ, Noordwijk, The Netherlands
}

\offprints{E. Maiorano, {\tt maiorano@bo.iasf.cnr.it}}
\date{Received 14 December 2004; Accepted 15 April 2005}

\abstract{ 
We report on the {\it BeppoSAX} data analysis of the afterglow of
Gamma--Ray Burst (GRB) 990123, one of the brightest GRBs detected by {\it
BeppoSAX}. 
Mainly due to its exceptional brightness, this is the only source for
which the Wide Field Cameras have allowed an early detection of the X--ray
afterglow between $\sim$20 and 60 min after the GRB trigger. Besides,
again for the first time, high-energy emission from the afterglow was
detected up to 60 keV. For the X--ray afterglow we found a power-law decay
with index $\alpha$ = 1.46 $\pm$ 0.04; the spectrum has a power-law shape
with photon index $\Gamma \sim 1.9$. The backward extrapolation of the
afterglow decay smoothly reconnects with the late GRB emission, thus
suggesting that both emissions are produced by the same phenomenon. An
extensive set of multiwavelength observations for the GRB 990123
afterglow made during the {\it BeppoSAX} pointing was collected from the
literature. The hard X--ray to radio range coverage allowed to construct a
spectral flux distribution and to perform an analysis of the GRB afterglow
in the context of the ``fireball" model. We also report the results of
temporal and spectral analysis of an X--ray source serendipitously
observed about 22$'$ north of the GRB afterglow, along with the optical
spectroscopy of its possible counterpart to establish the nature of this
source.
 
}

\maketitle

\section{Introduction}

The Gamma-Ray Burst (GRB) 990123 triggered the {\it BeppoSAX} (Boella et
al. 1997a) Gamma Ray Burst Monitor (GRBM) on 1999 Jan 23.4078 UT (Piro
1999a) and was simultaneously detected near the center of the field of
view of Wide Field Camera (WFC) no. 1 (Jager et al. 1997), with a
localization uncertainty of 2$'$ (error circle radius), at coordinates
(J2000) RA = 15$^{\rm h}$ 25$^{\rm m}$ 29$\fs$00, Dec = +44$^{\circ}$
45$'$ 00$\farcs$5 (Feroci et al. 1999). This burst was also seen by the
Burst And Transient Source Experiment (BATSE) on board the
Compton Gamma Ray Observatory ({\it CGRO}) starting approximately 18
s before the GRBM trigger; the T$_{90}$ duration of the burst, as measured
in 50--300 keV range, was 63.30 $\pm$ 0.26 s (Briggs et al. 1999). In
response to the trigger from BATSE, the Robotic Optical Transient Search
Experiment (ROTSE) at Los Alamos National Laboratories started wide angle
imaging of the GRB field and the first useful image was taken 22 s after
the onset of the burst (Akerlof et al. 1999). Thanks to the precise WFC
position promptly distributed (Feroci et al. 1999), a bright and rapidly
varying optical transient (OT) was found in the ROTSE images. The OT
peaked at magnitude $V \sim$ 9 about 50 s after the GRB onset and its
spectrum, obtained on 24.25 Jan 1999 UT, showed an absorption system at a
redshift $z$ = 1.600 (Kulkarni et al. 1999a; Andersen et al. 1999).

We refer the reader to the paper of Corsi et al. (2005) for a complete
study of the time-averaged and time resolved spectral properties of the
prompt emission and of its multi-wavelength optical/X--ray Spectral Flux
Distribution (SFD) analysis. Here we report a detailed analysis of the
X--ray prompt event data, in order to study the comparison between its
behaviour and the X--ray afterglow observations, and we provide the
complete analysis of the multiwavelength afterglow emission.

The paper is organized as follows: Sect. 2 describes the {\it BeppoSAX}
observations and the data analysis; the results are reported in Sect. 3
and the multiwavelength spectra are shown in Sect. 4. Discussion and
conclusions are presented in Sect. 5. Finally, in the Appendix A we
present X--ray/optical observations of a source serendipitously observed
during the {\it BeppoSAX} Narrow Field Instruments (NFIs) pointing of the
GRB 990123 afterglow.

When not otherwise indicated, errors for X--ray spectral parameters will
be reported at 90\% confidence level ($\Delta\chi^2$ = 2.7 for one
parameter fit), while errors for other parameters will be at 1$\sigma$;
upper limits will be given at 3$\sigma$.

\section{Afterglow observations and data reduction}

\subsection{WFCs data}

\begin{figure*}
\parbox{13.5cm}{
\psfig{file=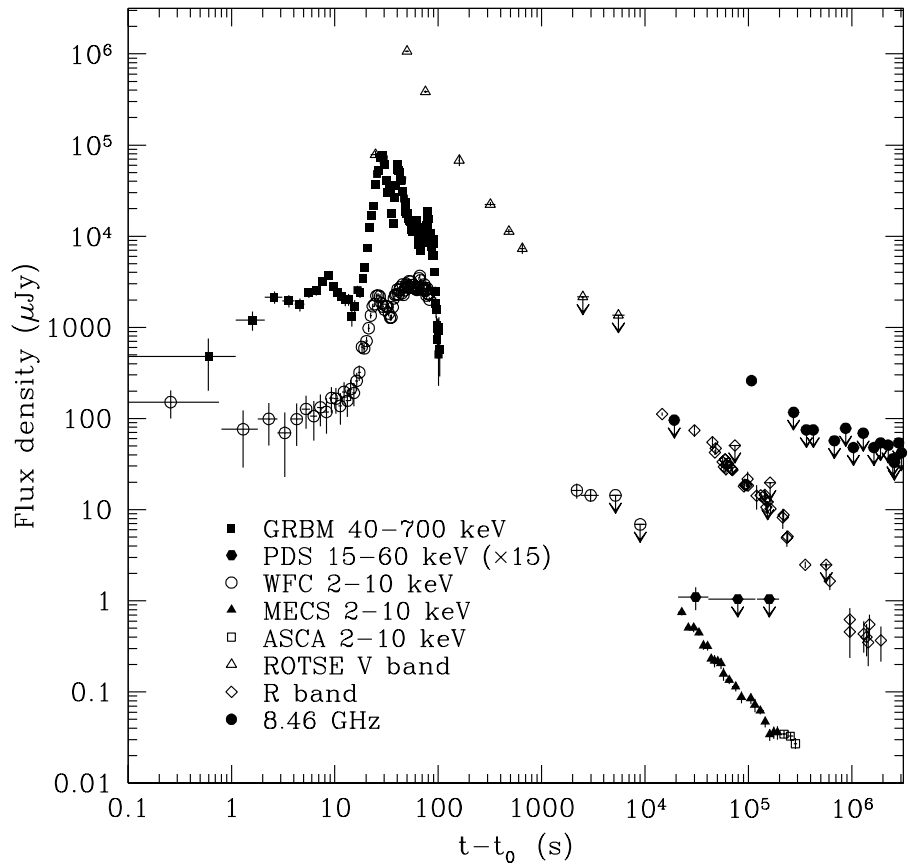,width=13cm}
}
\hspace{-1cm}
\parbox{6cm}{
\vspace{8.5cm}
\caption{Multiwavelength light curves of GRB 990123 and its afterglow;
$t_0$ corresponds to the time of the GRB onset.}
}
\end{figure*}

GRB 990123 was detected for about 100 s both in the GRBM 40--700 keV
and WFC 2--28 keV energy bands, showing two major pulses (Fig. 1).

The WFC instrument (Jager et al. 1997) consisted of two coded aperture
cameras, with a field of view of $40^{\circ}\times40^{\circ}$ and an
angular resolution of about 5$'$. The actual bandpass was 2--28 keV. The
axes of two GRBM units were co-aligned with those of the WFCs. Light
curves and spectra in the range 2--700 keV for GRB 990123 were then
extracted from the data acquired by WFC and GRBM.

The WFC data of the prompt event have been retrieved from the {\it
BeppoSAX} archive and analyzed by means of a standard package which
includes the "iterative removal of source" procedures (IROS, V. 105.108,
e.g. Jager et al. 1997), which implicitly subtract the background and the
other sources in the field of view.

Close to the end of the GRB the WFC no. 1 was pointing to the Earth
horizon. Atmospheric absorption then played an important role, as it
affected particularly the soft X--ray measurements close to the end of
the observation. At 80 s after trigger the Earth atmospheric absorption
was about 30\% at 5 keV and the subsequent decay of the 2--10 keV
X--ray light curve is substantially affected by the atmosphere.
Therefore, only the first 80 s of WFC data were plotted in Fig. 1. The
10--28 keV and 40--700 keV light curves are instead not significantly
influenced by the Earth atmosphere.

Given the high intensity of GRB 990123, we analyzed the WFC data
once the GRB position emerged again from the Earth occultation.  
Before the follow-up with the NFIs, the WFCs did not detect any
significant emission above 10 keV, but a fading X--ray source in the 2--10
keV band.

This is the first measurement of the X--ray afterglow in the time interval
around 30 min after the burst. We note that the late WFC emission cannot
be due to source serendipitously detected in the MECS because it is not 
positionally consistent with it and is two orders of magnitude brighter 
than the average 2--10 keV flux of RXS (see Appendix A).

\subsection{NFIs data}

The {\it BeppoSAX} NFIs included the following imaging telescopes: the
Low-Energy Concentrator Spectrometer (LECS, 0.1--10 keV; Parmar et al.
1997) and two Medium-Energy Concentrator Spectrometers (MECS, 1.5--10 keV;
Boella et al. 1997b). It also carried two non-imaging instruments based on
rocking collimators technique: the High Pressure Gas Scintillation
Proportional Counter (HPGSPC, 4--120 keV; Manzo et al. 1997) and the
Phoswich Detection System (PDS, 15--300 keV; Frontera et al. 1997); these
two instruments covered a field of view of 1$^{\circ}$ and 1\fdg3 FWHM,
respectively. Good NFI data were selected from intervals outside the South
Atlantic Geomagnetic Anomaly when the elevation angle above the
Earth limb was $>$5$^{\circ}$, when the instrument functioning was
nominal and, for LECS events, during spacecraft night time. The SAXDAS
2.0.0 data analysis package (Lammers 1997) was used for the extraction and
the processing of LECS, MECS and HPGSPC data. The PDS data reduction was
instead performed using XAS version 2.1 (Chiappetti \& Dal Fiume 1997).
The NFI observations were carried out in Target of Opportunity (ToO) mode
and started approximately 6 hrs after the trigger and continued during the
two following days. The on-source times for the two ToOs are 15.5 ks and
12.7 ks for the LECS, 45.4 ks and 36.6 ks for the MECS, 21.9 ks and 17.4
ks for the PDS respectively.

Background subtraction for LECS and MECS was performed using standard
library files while the background for the HPGSPC and for the PDS data was
evaluated from fields observed during off-source pointing intervals.  
Spectral and temporal analysis were also performed by using local
backgrounds extracted from the LECS and MECS event files: we found
consistent results with those obtained using standard background files.

A previously unknown X--ray source fully inside the WFC error box, 1SAX
J1525.5+4446, was detected almost at the center of the LECS and MECS
detectors, at the position (J2000) RA = 15$^{\rm h}$ 25$^{\rm m}$ 30$\fs$8
and DEC = +44$^{\circ}$ 46$'$ 21$\farcs$3 with a localization uncertainty
of 1$'$ along both coordinates. Based on the positional consistency with
the GRB position and the fading behaviour, similar to that of known GRB
afterglows, the source was identified as the X--ray afterglow of GRB
990123 (Piro 1999b). The high average intensity (2--10 keV) made it the
brightest X--ray afterglow detected by {\it BeppoSAX}.

Afterglow spectra and light curves for the LECS and the MECS during ToO1,
when the source was brighter, were extracted from 4$'$ and 3$'$ radius
regions, centered at the source position, respectively. For ToO2, the data
of both instruments were extracted from 2$'$ radius regions to reduce
background contamination, as the afterglow source was substantially
fainter during the second ToO. The data from the two MECS units were
summed together before extraction.

In order to derive a statistically significant 2--10 keV temporal
behaviour of the afterglow, we accumulated the light curve using a
variable binning along the two ToOs: bins of 3500 s for the first 40 ks of
ToO1, bins of 10000 s for the remaining 56 ks of ToO1, and bins of 15000 s
for ToO2.  The MECS 2--10 keV light curve of the afterglow is shown in
Fig. 1 and Fig. 2 (top panel), where the time origin is the GRB onset. The
average MECS spectrum (see Sect. 3.2) was used to convert the afterglow
light curve from counts s$^{-1}$ to erg cm$^{-2}$ s$^{-1}$.

During the first ToO a rapidly fading source was also detected by the PDS
in the 15--60 keV band. In the second part of ToO1 and in ToO2 the source
is not detected. This behaviour suggested that this was the high-energy
afterglow of the GRB 990123. Thus, thanks to the sensitivity of the
PDS, for the first time a GRB afterglow was also detected up to 60 keV.

In the MECS field another X--ray source located at about 22$'$ from the
X--ray afterglow source position at the coordinates RA = 15$^{\rm h}$
25$^{\rm m}$ 25$\fs$4 and Dec = +45$^{\circ}$ 07$'$ 06$''$, with 1$'$
error on both coordinates, was also well detected. This X--ray source is
positionally consistent with the {\it ROSAT} all-sky survey bright source
1RXS J152525.4+450706 (Voges et al. 1999), located at coordinates (J2000)
RA = 15$^{\rm h}$ 25$^{\rm m}$ 25$\fs$4, Dec = +45$^{\circ}$ 07$'$ 06$''$,
with an error radius of 10$''$. Hereafter we will refer to this source as
RXS. This object was not observed by the LECS as it was outside its field
of view. As for this source, we extracted its MECS data from a 3$'$ radius
region. The 15--28 keV data obtained with the PDS were corrected from the
contamination of RXS by assuming for this source a spectrum as described
in the Appendix A.

In the following we will assume an N$_{\rm H}$ Galactic value (Dickey \&
Lockman 1990) of 1.98$\times$10$^{20}$ cm$^{-2}$ along the GRB direction
and of 1.89$\times$10$^{20}$ cm$^{-2}$ along the RXS direction. The
N$_{\rm H}$ was modeled using the Wisconsin cross sections as implemented
in {\sc xspec} (Morrison \& McCammon 1983) and with solar abundances as
given by Anders \& Grevesse (1989).

The spectral analysis of the afterglow data (in Sect. 3.2) was performed
using the package {\sc xspec} (Dorman \& Arnaud 2001) v11.2.0. For the
spectral fitting, normalization factors were applied to the LECS and PDS
spectra following the cross-calibration tests between these instruments
and the MECS (Fiore et al. 1999).

\section{Results}

\subsection{The multiwavelength light curves}

Fig. 1 reports the $\gamma$--ray (40--700 keV), X--ray (2--10 keV and
15--60 keV), optical $V$ (Akerlof et al. 1999), and $R$ band
(Castro-Tirado et al. 1999; Galama et al. 1999; Kulkarni et al. 1999a;
Fruchter et al. 1999), and radio (8.46 GHz, Kulkarni et al. 1999b) light
curves of the prompt and afterglow emissions of GRB 990123. Time is
given in seconds since the GRBM trigger. The {\it ASCA} data points
are from Yonetoku et al. (2000), this observation started 57.4 hrs after
the trigger and lasted 80 ks. The temporal decay index of about 1.4 and
the photon spectral index $\Gamma$ of about 1.8 shown in Yonetoku et al.
(2000) are consistent with the values we obtained. Gunn $r$ magnitudes
were converted into Johnson-Cousins $R$ values by assuming $r - R = 0.45$
(Fruchter et al. 1999). $V$ and $R$ data were corrected for the Galactic
foreground reddening assuming $E(B-V)$ = 0.016 (Schlegel et al. 1998) and
they were converted into flux densities assuming the normalizations of
Fukugita et al. (1995). The host galaxy emission ($r = 24.5 \pm 0.2$;
Bloom et al. 1999) was subtracted from the $R$ data after conversion into
the Johnson-Cousins $R$ system as described above.

In Fig. 1 we plotted also the WFC data once the GRB position emerged again
from the Earth occultation. Although the data S/N was not good
enough to obtain meaningful spectra, this detection shows that the source
was still substantially active in X--rays about 2000 s after the prompt
event, at a flux of (2.84$\pm$0.54)$\times$10$^{-10}$ erg cm$^{-2}$
s$^{-1}$ assuming a Crab-like spectrum and then faded below
1.2$\times$10$^{-10}$ erg cm$^{-2}$ s$^{-1}$ about 2 hrs later.

Using a single power-law model to describe the temporal decay ($F_{\rm
X}(t) \propto t^{-\alpha}$), the 2--10 keV MECS measurements are well fit
($\chi^2$/dof = 19.4/19) with an index $\alpha$ = 1.46$\pm$0.04 (dashed
line in Fig. 2, top panel). The {\it BeppoSAX} time coverage stops
across the break observed in the optical $R$-band light curve
$\sim$50 hrs after the GRB (Fruchter et al. 1999; Kulkarni et al. 1999a,
Castro-Tirado et al. 1999). The low statistics in the X--ray data
obtained after the optical break time does not allow to understand whether
a slope change, simultaneous with the optical break, occurs. Actually, the
{\it ASCA} data acquired after the end of the NFI observations (Fig. 1)
show a deviation, from the trend seen with {\it BeppoSAX}, in the form of
an excess of emission rather than of a break. A possible interpretation to
this behaviour will be given in Sect. 5.

For the LECS and the PDS data a temporal analysis similar to that obtained
with the MECS was not possible because of the lower statistics. However,
we do not expect substantial differences as the spectral X--ray evolution
is consistent with being achromatic (see Sect. 3.2).

\begin{figure}
\psfig{file=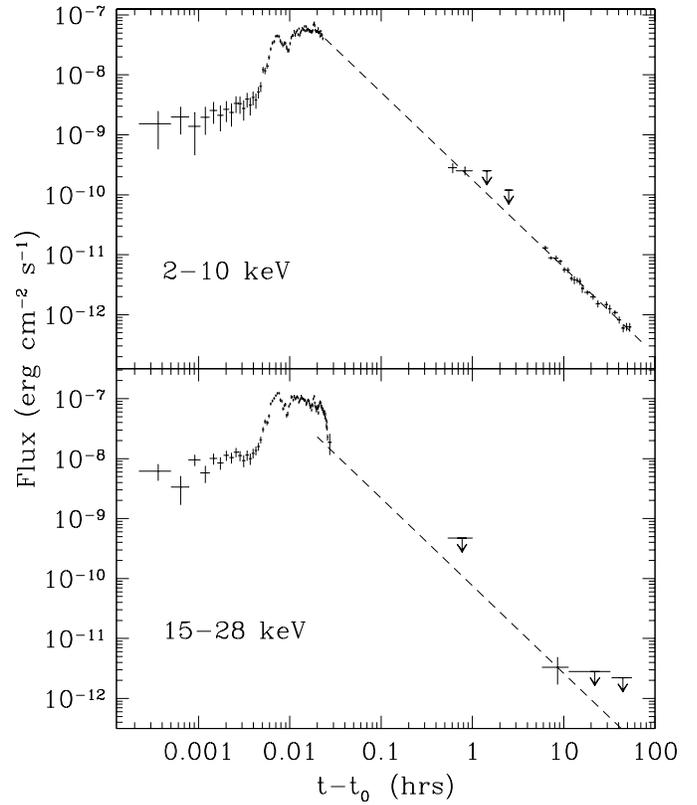,width=9cm,angle=0}
\caption{(Top panel) 2--10 keV light curve constructed using WFC and MECS
data. The dashed line is the best-fit decay obtained from the X--ray
afterglow data. As can be seen, the extrapolation reconnects smoothly with
the late-time WFC data points and upper limits, suggesting that the X--ray
afterglow had already started $\sim$30 min after the GRB 990123
prompt event. (Bottom panel) Light curve in the energy range 15--28 keV
constructed using WFC and PDS data. The dashed line, with slope equal to
the one determined from the 2-10 keV afterglow data and passing through
the PDS detection, smoothly reconnects with the late prompt event data
points.}
\end{figure}

It is the first time in which an X--ray afterglow is detected above 10
keV. In the 15--28 keV energy range, starting from about 2000 s after the
GRB, we derived the light curve of GRB 990123 using the WFCs data
for the prompt emission and for the first time the PDS data.

Fig. 2 (top and bottom panels) shows the results. Concerning the prompt
event we considered the WFC data until the atmospheric absorption was
negligible (that is, larger than 30\%, i.e. up to 80 s after the onset of
the burst). In both figures the dashed line indicates the afterglow
best-fit power-law decay with the slope $\alpha$ = 1.46 $\pm$ 0.04 found
with the 2--10 keV MECS data.

As can be seen from Fig. 2, the backwards extrapolation of the 2--10 keV
(top panel) afterglow best-fit decay slope is fully consistent with the
late WFC data points and upper limits, and reconnects with the last data
points of the prompt event. The same extrapolation was also done in the
15--28 keV light curve (bottom panel) by normalizing the power-law decay
to the single PDS detection. Again, we find that it is fully consistent
with the last data from the prompt event and with the WFC upper limit
obtained $\sim$30 min after the GRB onset. This behaviour in both
energy ranges suggests that the last WFC points already represent the
afterglow emission. Thus, mainly due to its exceptional brightness, this
is the only source for which the Wide Field Cameras have allowed an early
detection of the X--ray afterglow between $\sim$20 and 60 min after the
GRB trigger.

\subsection{X--ray afterglow spectral analysis}

\begin{table*}[th!]
\caption[] {Results of the time-resolved spectral fits of the GRB 
990123 afterglow {\it BeppoSAX} NFI observations. In each case the 
N$_{\rm H}$ amount (in squared parentheses) was fixed at the Galactic value.} 
\begin{center}
\begin{tabular}{cccccc}
\noalign{\smallskip}
\hline
\hline
\noalign{\smallskip}
ToO & Start time & duration & $\Gamma$ & N$_H$ & $\chi^2$/dof  \\
  & (Jan 1999 UT) & (ks) &  & (10$^{20}$ cm$^{-2}$) &  \\
\hline
\hline
\noalign{\smallskip}
 1 (1$^{\rm st}$ part) & 23.6495 & 20 & $1.94^{+0.12}_{-0.13}$ & 
	[1.98] & 40/50  \\
 1 (2$^{\rm nd}$ part) & 23.8810 & 76.2 & $2.07^{+0.11}_{-0.12}$ & 
	[1.98] & 56/54  \\
 2 & 24.8132 & 76.5 & $1.86^{+0.29}_{-0.29}$ & [1.98] & 16/16  \\
\noalign{\smallskip}
\hline
\hline
\end{tabular}
\end{center}
\end{table*}

The afterglow spectrum, due to the high brightness of the source, has a
very good statistical quality (as, for example, the X--ray afterglow of
GRB010222 observed by {\it BeppoSAX}; in 't Zand et al. 2001). In
addition, for the first time a GRB X--ray afterglow was detected up to 60
keV during the first 20 ks of the {\it BeppoSAX} NFI pointing.

As remarked earlier, since the MECS images include the source RXS, we
first estimated its contamination to the afterglow in the PDS data. To
this aim, given that RXS appears to be a steady X--ray source during the
two {\it BeppoSAX} ToOs, we extracted for it a time-averaged MECS spectrum
accumulated over the two ToOs to increase the S/N. The 2--10 keV best-fit
model of the RXS spectrum (described in the Appendix A) was then
extrapolated to the 15--60 keV range, and a flux value was derived for it.
Considering its offset position with respect to the PDS field of view and
taking into account the triangular response of this instrument (Frontera
et al. 1997), the actual contribution of RXS is
1.2$\times$10$^{-12}$ erg cm$^{-2}$ s$^{-1}$ between 15 and 60 keV, which
corresponds to $\sim$12\% of the total flux detected in the 15--60 keV
band during the first 20 ks of ToO1. Although this contribution is
comparable to the statistical errors in the PDS, it affected
systematically the PDS flux from GRB 990123, therefore we have
subtracted it.

In order to test the presence or not of a spectral evolution, the entire
observation was divided into three parts and time-resolved spectra were
obtained. A simultaneous combined LECS, MECS and PDS spectrum was obtained
in the first 20 ks of the observation, between Jan 23.6495 and Jan
23.8810, when the source was brightest and well detected in all of these
instruments (Fig. 3). We fitted this spectrum with a photoelectrically
absorbed power-law. As the fitted value of N$_{\rm H}$,
(0.9$^{+15}_{-0.9}$)$\times$10$^{20}$ cm$^{-2}$, is consistent with the
Galactic column density along the GRB direction, we fixed its value to the
Galactic one. We then found for the best-fit photon index the value
$\Gamma$ = $1.94^{+0.12}_{-0.13}$. We note that the N$_{\rm H}$ value
we found letting this parameter free to vary in the fit is consistent with
that obtained by Yonetoku et al. (2000) from the {\it ASCA} observation
carried out after the end of the {\it BeppoSAX} pointing.

In the following 76.2 ks of ToO1, between Jan 23.8810 and Jan 24.7629, and
during ToO2 (i.e. between Jan 24.8132 and Jan 25.6986), the afterglow was
detected in the 0.6--10 keV range only. The spectra accumulated over these
two time intervals were also analyzed, and the fits were made assuming
again a power-law description; the best-fit photon indices are consistent
with that obtained from the first 20 ks spectrum (see Table 1). Therefore
we can say that the three time resolved spectra are consistent with no
spectral variation, that is, the X--ray afterglow evolution is achromatic.
Unfortunately the low statistics of the data between the optical
break epoch (2 days after the GRB) and the end of the {\it BeppoSAX}
observations (2.29 days after the GRB) does not allow us to perform a
sensible spectral analysis in this time interval.

Assuming the best-fit spectral model obtained in the first 20 ks data, we
derived an upper limit of 7$\times$10$^{-12}$ erg cm$^{-2}$ s$^{-1}$ to
the PDS afterglow flux in the 15--60 keV band during both time intervals 
described above: this is consistent with the extrapolation of the 
0.6--10 keV spectrum.

\begin{figure}
\psfig{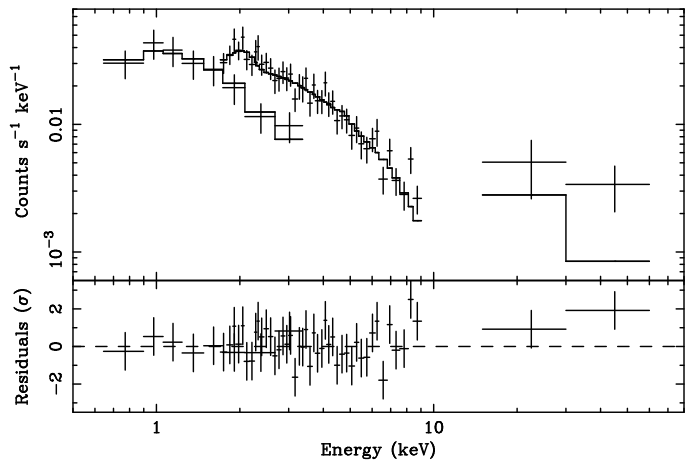}
\caption{The 0.6--60 keV afterglow of GRB 990123 during the first 20 
ks of the {\it BeppoSAX} ToO1. An absorbed power-law with photon index 
$\Gamma = 1.94$ best fits the data.} 
\end{figure}

We note that the LECS-MECS inter-calibration factor we obtained from the
fits (i.e., $\sim$0.6) was low compared to the range (0.7--1.0) obtained
from the cross-calibration tests between the LECS and MECS (Fiore et al.
1999). In order to investigate this unusual behaviour, as also explored by
Stratta et al. (2004), we assumed in the fit an additional N$_{\rm H}$
reproducing further absorption local to the GRB host at $z$ = 1.6, but in
this case the N$_{\rm H}$ value of the additional local absorption was
unconstrained and we only obtained a 90\% upper limit of
1.4$\times$10$^{22}$ cm$^{-2}$ for it. The difference between the results
of Stratta et al. (2004) and ours can be explained by the different energy
band and integration intervals considered. Anyway, we point out that
Stratta et al. (2004) infer the significance of the local absorption on
the basis of the usual F--test for the addition of a further
component but this test, as discussed by Protassov et al. (2002),
cannot be used in this case.

This deficit of the LECS normalization factor could also possibly be due
to a non-perfect simultaneity of LECS and MECS observations, that is, LECS
lost a fraction of the decaying afterglow emission due to its
observational constraints. In order to check this hypothesis, we extracted
simultaneous LECS and MECS spectra using the same time windows. However,
again, the fit to these spectra (which have a lower S/N) returns a
LECS-MECS inter-calibration factor consistent with 0.6. Another possible
explanation for the discrepancy is the presence of a spectral break around
$\sim$ 0.5 keV (see Sect. 5).

\section{The broadband spectrum of the afterglow}

Since during the entire NFI observation, the optical and near-infrared
(NIR) light curves did not show chromatic evolution, we considered the SFD
already presented by Galama et al. (1999) and we completed it with the NFI
X--ray data. We referred all data points to the date 24.65 Jan 1999 UT.

The optical flux densities at the wavelengths of $UBVRIHK$ bands have been
derived without subtracting any host galaxy contribution, because it was
negligible at the epoch we selected (cfr. Fruchter et al. 1999). We
assumed $E(B-V)$ = 0.016 to deredden the data, and the normalizations
given in Fukugita et al. (1995) and Bersanelli et al. (1991) to obtain the
corresponding flux densities in optical and NIR, respectively. When
needed, we rescaled the data to the corresponding reference date using the
optical power-law decay with index $\alpha_{\rm opt}$ = 1.10 $\pm$ 0.03
(Kulkarni et al. 1999a). We rescaled the flux of the broadband X--ray
spectrum (0.6--60 keV) using the power law decay measured from 2--10 keV
data ($\alpha_{\rm X}$ = 1.46; see Sect. 3.1). The optical and NIR
extinction-corrected data, together with the 0.6--60 keV spectrum observed
at 24.65 Jan 1999 UT, are shown in Fig. 4. In this figure we also included
radio data and upper limits as reported in Galama et al. (1999).

Since the optical and X--ray band light curves showed different temporal
decays, we independently fitted with a power law the optical/NIR spectrum
and we obtained an energy spectral index value of $\beta_{\rm opt}$ = 0.60
$\pm$ 0.04 (with $\chi^2$/dof = 2.3/5), flatter than the X--ray one
($\beta_X$ = 0.94 $\pm$ 0.07 at 1$\sigma$). The presence of a spectral
turnover between optical and X--ray bands could be identified with the
change of spectral slope (i.e., a steepening) at a frequency which we
identified with the cooling frequency $\nu_{\rm c}$ in the framework of
the synchrotron fireball model (Sari et al. 1998).

Assuming a negligible host absorption and using the optical/NIR and X--ray
slopes, we obtained for $\nu_{\rm c}$ the value 1.1$\times$10$^{17}$ Hz,
corresponding to 0.5 keV, i.e. the lower-energy edge of the X--ray energy
band covered by {\it BeppoSAX}. This bend around 0.5 keV can be a possible 
explanation for the deficit found in the LECS-MECS inter-calibration 
constant.

\begin{figure}
\psfig{file=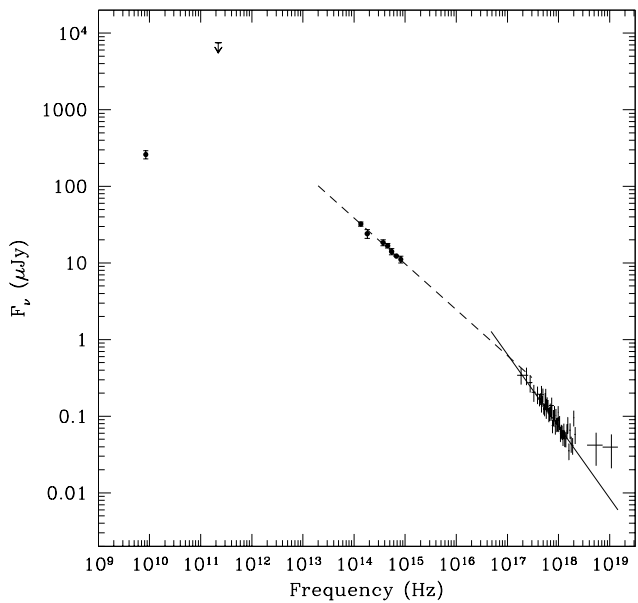,width=9cm}
\caption{The SFD of GRB 990123 at 24.65 Jan 1999 UT. The optical,
NIR and radio data presented here are from Galama et al. (1999). As
regards the radio data, only those acquired strictly within 0.03 d of 
the reference epoch above are included in the plot. The dashed line is
the best-fit power law (with $\beta_{\rm opt}$ = 0.60) describing optical
and NIR data, while the solid line is the power law (with $\beta_{\rm X}$
= 0.94) which best fits the {\it BeppoSAX} NFI data.}
\end{figure}

\section{Discussion}

GRB 990123 was a very considerable astrophysical event: it was one
of the brightest GRBs ever detected with {\it BeppoSAX}, and the only one
up to now for which prompt optical emission simultaneously with the
high-energy one was detected (Akerlof et al. 1999). Here we report two
other records from this GRB: (i) for the first time an X--ray afterglow is
detected starting 2000 s after the prompt event, and (ii) for the first
time emission up to 60 keV is detected from a GRB afterglow.

Many properties of GRB 990123, like the X--ray afterglow fluence
with respect to that of the prompt emission or with respect to the
gamma-ray fluence, are similar to those of other GRBs. Following Frontera
et al. (2000), the estimated 2--10 keV total afterglow fluence from 60 s
(corresponding to $\sim$60\% of the GRB time duration) to $1 \times
10^{6}$ s since the GRB start, is given by $S_a = 9.0 \times 10^{-6}$ erg
cm$^{-2}$, which is about 4.5\% of the prompt fluence in the 40--700 keV
band ($S_\gamma$ = $2.0 \times 10^{-4}$ erg cm$^{-2}$; Corsi et al. 2005).
On the other hand, the afterglow-to-prompt fluence ratio in the
2--10 keV energy range is $\sim$2.9 (the 2--10 keV prompt emission fluence
is $S_{\rm X}$ = $3.1 \times 10^{-6}$ erg cm$^{-2}$, Corsi et al. 2005).
In fact this ratio is an upper limit because of the absorption due to the
Earth atmosphere in the late part of the prompt event. In any case these
values are well within the observed range of values found GRBs by Frontera
et al. (2000) for a sample of GRBs.

The X--ray time emission history also is similar to other GRB events.
Figure 2 shows quite a smooth connection between the prompt event and the
afterglow. This was already suggested by Frontera et al. (2004) from the
analysis of a sample of GRBs observed with {\it BeppoSAX}. However, thanks
to the brightness of the event, for the first time the WFCs were able to
detect the GRB until $\sim$1 hr after the trigger, allowing us to
sample the burst behaviour during an unexplored temporal range from the
end of the prompt emission until the starting time of the ToO observations
(6 hrs after the event). We find that in this time interval the X--ray
emission is consistent with being due to GRB afterglow.

The more interesting properties concern the multiwavelength spectrum and
fading law of the afterglow emission. The 2--10 keV afterglow fading law,
from the end of the prompt emission up to $\sim 2$ days is well
described by a power-law with index $\alpha_X = 1.46 \pm 0.04$. On the
other hand, from about 3.5 hrs from the burst until about 2 days the
optical afterglow in the $R$-band shows a power-law decay with index
$\alpha_{\rm opt}$ = 1.10 $\pm$ 0.03 (Kulkarni et al. 1999a; Fruchter et
al. 1999). After this epoch, the X--ray light curve seems to flatten
rather than to show a break like that in the optical; then, it resumes the
decay. The statistics is however not good enough to tell whether this
decay is similar to that seen in the optical after the break.
 
For what concerns the multiwavelength spectrum, as showed in Sect. 4, we
expand and improve the work made on the SFD by Galama et al. (1999) thanks
to information obtained from the X--ray spectrum. We find that the
spectrum in the optical/NIR band and in the X--ray bands is well described
by a power-law with index, $\beta_{\rm opt} = 0.60 \pm 0.04$ in the former
band, while $\beta_{\rm X} = 0.94 \pm 0.07$ in the latter one. From these
spectral slopes we derive a cooling frequency of $\nu_{\rm c} \sim$
1.1$\times$10$^{17}$ Hz corresponding to about 0.5 keV.

The comparison of the spectral and temporal results is intriguing. In the
context of the fireball model, assuming an isotropically adiabatic
expansion of the fireball into a uniform interstellar medium (Sari et al.
1998), for a synchrotron emission the difference $\Delta\alpha_e$ between
the optical and X--ray decay index is expected to be $\alpha_{\rm X} - 
\alpha_{\rm opt} = (3p-2)/4 - 3(p-1)/4 = 1/4$ where $p$ is the electron
energy distribution index. Indeed we find that the measured difference
$\Delta\alpha_m$ is $0.36 \pm 0.05$ which is consistent with the expected
$\Delta\alpha_e$. Also the expected difference between the optical and
X--ray spectral slope indices $\Delta\beta_e$ = $\beta_{\rm X} -
\beta_{\rm opt} = p/2 - (p-1)/2 = 1/2$ that is fully consistent with the
measured difference $\Delta\beta_m = 0.34 \pm 0.08$. However, from the
above relations also a link between temporal and spectral properties is
expected: for frequencies below $\nu_{\rm c}$, i.e. in the optical band,
$\beta_{\rm opt} = 2/3 \alpha_{\rm opt}$ while for frequencies above 
$\nu_{\rm c}$, i.e. in the X--ray band, $\beta_{\rm X} = 2/3 \alpha_{\rm X} 
+ 1/3$. We find that these relations are not satisfied by the observational 
data either in the optical band or in the X--ray band. Indeed 
$\beta_{\rm opt}/\alpha_{\rm opt} = 0.54 \pm 0.04$ against an expected value 
of 0.66, while $(\beta_{\rm X} - 1/2)/\alpha_{\rm X} = 0.30 \pm 0.05$ 
against still a value of 0.66. As can be seen, both measured ratios are 
statistically inconsistent with the expectations.

If we take into account the local absorption ($A_V\approx 1.1$, Savaglio
et al. 2003), $\beta_{\rm opt}$ should be even lower and
consequently the discrepancy between the expected and the measured ratio
$\beta_{\rm opt}/\alpha_{\rm opt}$ becomes even higher. Similarly,
assuming a gray absorption (e.g., Maiolino et al. 2001), would not
alleviate the mentioned discrepancy as it would influence only the
power-law normalization of the optical/NIR spectrum moving $\nu_{\rm c}$
toward the optical band.

Therefore, a self-consistent interpretation of the afterglow with pure
syncrotron emission is not viable. A partial solution is presented by
Corsi et al. (2005) who envisage the presence of an Inverse Compton (IC)
component in the X--ray band. The emergence of this IC component
$\sim 2$ days after the GRB could explain the emission excess detected at
the end of the {\it BeppoSAX} observations and during the first part of
the {\it ASCA} pointing (Fig. 1). However, as demonstrated above the
inconsistency not only concerns the X--ray band but also the optical band.
So a more complex model than an IC in the only X--ray band is required.

\begin{acknowledgements}
{\it BeppoSAX} was a joint program of Italian (ASI) and Dutch (NIVR) space 
agencies. This work has made use of the ASI Science Data Center Archive at 
ESA/ESRIN, Frascati (Italy), of the NASA's Astrophysics Data System and of 
the SIMBAD database, operated at CDS, Strasbourg (France). This research 
has been partially supported by ASI.
\end{acknowledgements}

\appendix

\section{The source RXS in the MECS field}

As explained above, in order to check for any possible contamination in
the PDS energy range (15--60 keV) due to the X--ray source detected in the
MECS field, we extracted and analyzed the light curve information on this
source as observed by {\it BeppoSAX}. The 2--10 keV light curve during the
entire {\it BeppoSAX} observation showed that it had a steady behaviour.
We therefore accumulated its spectrum averaged over both ToOs, and fitted
the spectral data making use of the available {\it BeppoSAX} off-axis
matrices. The averaged spectrum was satisfactorily fitted ($\chi^2$/dof =
23/21) using an absorbed power law, and we found for the best-fit
spectral index the value $\Gamma$ = 1.60$^{+0.26}_{-0.25}$ (see Fig. A.1).
This implies an unabsorbed 2--10 keV flux of 1.3$\times$10$^{-12}$ erg
cm$^{-2}$ s$^{-1}$ from this source.

Zickgraf et al. (2003) proposed as the possible optical counterpart to
this object a fairly bright QSO (source `2' in Fig. A.2, with no reported
redshift) located 19$''$ from the X--ray position, thus formally outside
the {\it ROSAT} error box. Actually, by looking at the
DSS-II-Red\footnote{available at\\
\texttt{http://archive.eso.org/dss/dss/}} image of the field in Fig. A.2,
a relatively fainter object (source `1') is found practically at the
center of the {\it ROSAT} error circle. We therefore started an optical
spectrophotometric campaign in order to understand which of the two
objects is the responsible of the X--ray emission.

\begin{figure}
\psfig{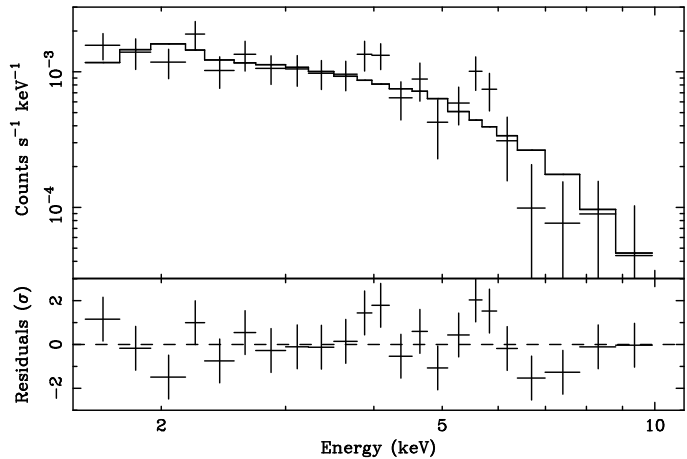}
\caption{MECS 1.8-10 keV spectrum of the source 1RXS J152525.4+470506, 
averaged over the two {\it BeppoSAX} pointings. Data are well fitted by
an absorbed power law with photon index $\Gamma = 1.60^{+0.26}_{-0.25}$.}
\end{figure}

\begin{figure}
\psfig{file=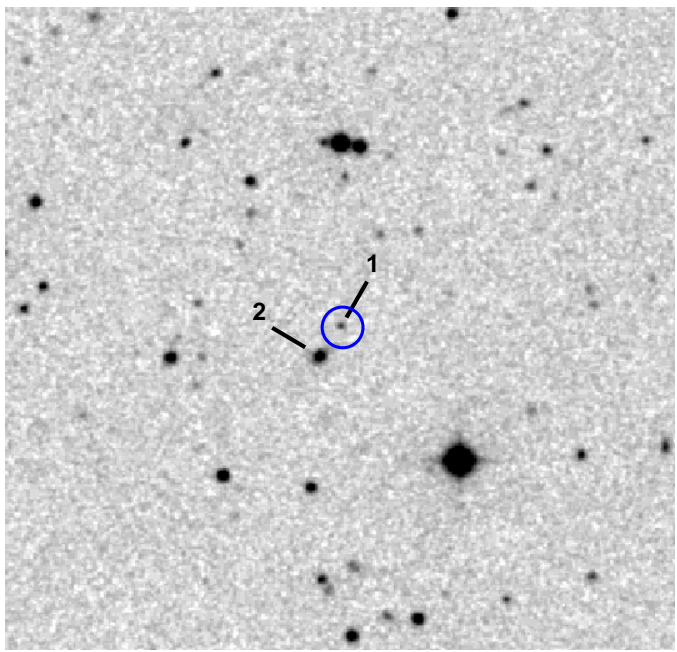,width=9.0cm}
\caption{DSS-II-Red image of the field around the source RXS. North is at
top, East is to the left. The field size is about 5$'$$\times$5$'$. The
{\it ROSAT} error circle for the source is also indicated. Only one
object, source `1', is present in the {\it ROSAT} uncertainty circle;
however, the counterpart proposed by Zickgraf et al.  (2003), source `2',
is the brighter object lying southeast of the {\it ROSAT} error circle and
just outside it.}
\end{figure}

\begin{table}
\caption[]{Results of the $UBVRI$ photometry acquired in Loiano on 1-3 
August 2003 on the two sources in Fig. A.2. Errors are given at 1$\sigma$ 
confidence level.}
\begin{center}
\begin{tabular}{rccc}
\noalign{\smallskip}
\hline
\hline
\noalign{\smallskip}
\multicolumn{1}{c}{Exposure} & Filter & Source `1' & Source `2' \\
\multicolumn{1}{c}{time (sec)} &  &  &  \\
\hline
\hline
\noalign{\smallskip}
 4$\times$1800 & $U$ & 21.0$\pm$0.1 & 16.20$\pm$0.02 \\
 4$\times$1200 & $B$ & 20.21$\pm$0.03 & 16.66$\pm$0.02 \\
  2$\times$900 & $V$ & 19.86$\pm$0.04 & 17.04$\pm$0.02 \\
  2$\times$600 & $R$ & 19.55$\pm$0.06 & 16.93$\pm$0.02 \\
  3$\times$600 & $I$ & 18.60$\pm$0.05 & 15.94$\pm$0.04 \\
\noalign{\smallskip}
\hline
\hline
\end{tabular}
\end{center}
\end{table}

To this aim, medium-resolution optical spectra (dispersion: 4
\AA/pix) of object `2' and optical $UBVRI$ observations of the field in
the Johnson-Cousins photometric system were performed in Loiano (Italy) on
27 April 2003 and 1-3 August 2003, respectively, with the 1.5m ``G.D.
Cassini" telescope plus BFOSC under fairly good weather conditions, with
seeing 1$\farcs$5. We also acquired a medium-resolution spectrum
(dispersion: 2 \AA/pix) of the fainter object with the WHT plus ISIS (red
arm) on 15 April 2003 under poor sky conditions (seeing: 2$\farcs$0, high
humidity and thick cirrus). All images were bias-subtracted and
flat-fielded following the standard procedure.

We chose standard Point Spread Function (PSF) fitting for all photometry
measures, as the objects of interest appeared point-like in our images,
analysis package PSF--fitting algorithm (Stetson 1987) running within
MIDAS\footnote {MIDAS (Munich Image Data Analysis System) is developed,
distributed and maintained by ESO (European Southern Observatory) and is
available at {\tt http://www.eso.org/projects/esomidas}}. Photometry was
flux-calibrated using the Landolt (1992) field PG 1633+099. The $UBVRI$
photometry results are shown in Table A.1. The errors associated with
these measurements represent statistical uncertainties (at 1$\sigma$)
obtained with the standard PSF--fitting procedure.

Spectra were background subtracted and optimally extracted (Horne 1986)
using IRAF\footnote{IRAF is the Image Analysis and Reduction Facility made
available to the astronomical community by the National Optical Astronomy
Observatories, which are operated by AURA, Inc., under contract with the
U.S. National Science Foundation. It is available at {\tt 
http://iraf.noao.edu/}}. The spectra acquired in Loiano and
with WHT were wavelength calibrated with Helium-Argon lamps and
Copper-Neon-Argon lamps, respectively. In both cases the spectroscopic
standard BD+33$^\circ$2642 (Oke 1990) was used to flux-calibrate the spectra.

Spectroscopy shows that, whereas we did not detect any characteristic
feature in the spectrum of source `1' because of the low S/N ratio, from
source `2' we observed several broad Balmer emissions and a number of
narrow emission lines which we identified with the forbidden transitions
of [O {\sc iii}] $\lambda\lambda$4959,5007 \AA~and [N{\sc ii}]
$\lambda\lambda$6548,6583 \AA. All these lines are consistent with having
a redshift $z$ = 0.253 $\pm$ 0.001.

The presence of broad redshifted Balmer lines suggests that source `2' is
a Type 1 AGN. Given that this kind of AGNs have an average X--ray spectral
index $\Gamma$ = 1.67 $\pm$ 0.14 (Mushotzky et al. 1993), the
identification of RXS with this optical object (Zickgraf et al. 2003) is
strengthened, even if the source `1' inside the {\it ROSAT} error circle
cannot be completely ruled out as counterpart.

\end{document}